# FPGA-Based Mini X-Ray Detector Front-End

Kris Paetow and D.G. Perera
Department of Electrical and Computer Engineering,
University of Colorado Colorado Springs
Colorado Springs, Colorado, USA

**Abstract**

Medical imaging systems require reliable front-end electronics that can acquire sensor data, process image information, identify errors, and communicate results to other parts of the system. In applications such as X-ray imaging, CT, PET, ultrasound, and other diagnostic imaging systems, the electronics must often handle large amounts of data while maintaining predictable timing and low-latency operation. Because of these requirements, FPGAs (Field Programmable Gate Arrays) are commonly useful for medical imaging and signal-processing applications. In this project, the medical imaging concept is simplified into a small FPGA-based frontend demonstration.

# 1 Introduction

## 1.1 Current Scenario

Medical imaging systems require reliable front-end electronics that can acquire sensor data, process image information, identify errors, and communicate results to other parts of the system. In applications such as X-ray imaging, CT, PET, ultrasound, and other diagnostic imaging systems, the electronics must often handle large amounts of data while maintaining predictable timing and low-latency operation. Because of these requirements, FPGAs (Field Programmable Gate Arrays) are commonly useful for medical imaging and signal-processing applications. FPGAs provide parallel processing, reconfigurable hardware logic, and the ability to implement custom data paths for real-time image processing tasks.

FPGA-based medical imaging research has shown that these devices can be used directly in front-end electronics. For example, Haselman et al. describe FPGA-based front-end electronics for positron emission tomography and note that modern FPGAs are capable of performing complex discrete signal-processing algorithms at high clock rates [1]. This is important because many medical imaging systems depend on front-end electronics that can acquire signals, process them, and pass the results forward without relying only on slower sequential software processing.

More broadly, FPGA design is useful in medical devices because it supports high computing power, low latency, reconfigurability, and faster design iteration compared with fixed custom hardware. KritiKal Solutions describes FPGAs as useful for medical devices because they can support rapid prototyping, custom task acceleration, low power operation, and post-manufacturing upgrades [2]. These same advantages are relevant to medical imaging, where the design may need to be tested, modified, or expanded before moving toward a final product. VLSI First also highlights that FPGA-based medical imaging systems can perform real-time image processing, filtering, image recognition, and signal-processing tasks, making them a good fit for systems that need both speed and flexibility [3].

In this project, the medical imaging concept is simplified into a small FPGA-based frontend demonstration. Instead of building a complete X-ray detector, the system uses a grayscale test image to represent detector pixel data. The FPGA then analyzes the image for offset and gain-related pixel errors, stores or reports error locations, applies correction logic, and returns the corrected image to the host computer. This makes the project a small-scale model of the type of digital processing that may occur between an imaging sensor and a final displayed image.

## 1.2 Problem Description

The problem addressed in this project is how to build and demonstrate a simplified FPGA front-end system for detecting and correcting image errors that are representative of issues found in detector-based imaging systems. In an X-ray detector, each pixel should ideally respond in a predictable way to the incoming signal. However, real detector systems can contain pixel-to-pixel variations. Two important examples are offset error and gain error. A high offset error causes a pixel to appear artificially brighter or darker because its baseline value is shifted. A high gain error causes a pixel to respond too strongly or too weakly compared with neighboring pixels because its sensitivity is different from normal.

The project is not intended to reproduce the full complexity of a commercial X-ray detector. Instead, it focuses on the digital logic concepts behind a detector front-end: data transfer, image storage, pixel analysis, error reporting, image correction, and system-level control. By reducing the problem to a small grayscale image and a few controlled error types, the project creates a practical FPGA demonstration that can be tested within the scope of the class.

## 1.3 Motivation

The motivation for this project comes from my work in an X-ray manufacturing lab. In that environment, I have experience with sensor design and a general understanding of the fabrication process used to build detector-related devices. However, the sensor itself is only one part of a complete imaging system. After the sensor produces electrical or digital data, additional electronics are required to read out the data, process it, correct errors, and prepare the image for use. Because of this, I wanted to use the final project as an opportunity to get a small glimpse into the programming and digital logic side of X-ray detector electronics. The project connects my existing interest in X-ray sensor manufacturing with the FPGA concepts covered in this course. It also provides a practical way to understand how hardware logic can be used for image transfer, analysis, correction, and feedback.

Another motivation is that FPGA prototyping is especially useful for systems that may need to change during development. Medical imaging hardware can require many design iterations before a final architecture is selected. An FPGA allows the logic to be modified without fabricating a new chip, which makes it useful for early-stage testing and proof-of-concept development. In this project, that same idea is applied at a smaller scale by using the FPGA as a reconfigurable platform for a mini image-processing front end.

## 1.4 Main Goal

The main goal of this project is to design and demonstrate a simplified FPGA-based mini X-ray detector front-end system that can receive a grayscale image from a host PC, store the image in BRAM, detect offset and gain-related pixel errors, display detected error locations on an LCD, apply basic correction logic, and return the corrected image to the PC. The expected result is a working proof-of-concept system that demonstrates the major steps of a detector frontend pipeline: image input, memory storage, error analysis, user feedback, correction, and image output. The project also uses FPGA board resources such as UART, BRAM, pushbuttons, LEDs, and an LCD display to show both the internal processing flow and the final corrected image result.

# 2 Detailed Project Description

## 2.1 Project Overview

In real digital radiography systems, detector correction is used to make the pixel array respond as uniformly as possible before the image is displayed. Offset and gain corrections are needed because pixels may differ in dark signal, sensitivity, and response to the same X-ray exposure, and the external readout electronics can also introduce gain and offset variation. These corrections are commonly handled using offset subtraction and flat-field correction, where calibration images are used to estimate each pixel's baseline error and response sensitivity. The goal is for all pixels to have effectively identical response curves, so that a uniform X-ray exposure produces a uniform image instead of a pattern caused by detector or electronics nonuniformity. Bad pixels or lines may also be identified during flat-field analysis and corrected using nearby pixel values, such as mean or median filtering [4].

This project implements a simplified version of that idea. Instead of using actual X-ray calibration frames, the system uses a generated grayscale test image with known artificial offset and gain errors. The system sends the grayscale image to the FPGA, where it is stored on the BRAM, errors are corrected, then the corrected image is sent back to the computer.

## 2.2 System-Level Design

Figure 1 shows the high level system diagram of the design. The concept of the design is straightforward: send a grayscale image created through Python from the PC to the FPGA, analyze and correct any "defective" pixels on the FPGA, then return the corrected image to the PC. The FPGA will also show the number of errors found on the LEDs, and the LCD will display the locations. The image is stored on the BRAM of the FPGA so that the image process module and the LCD can read the pixels and errors with ease.

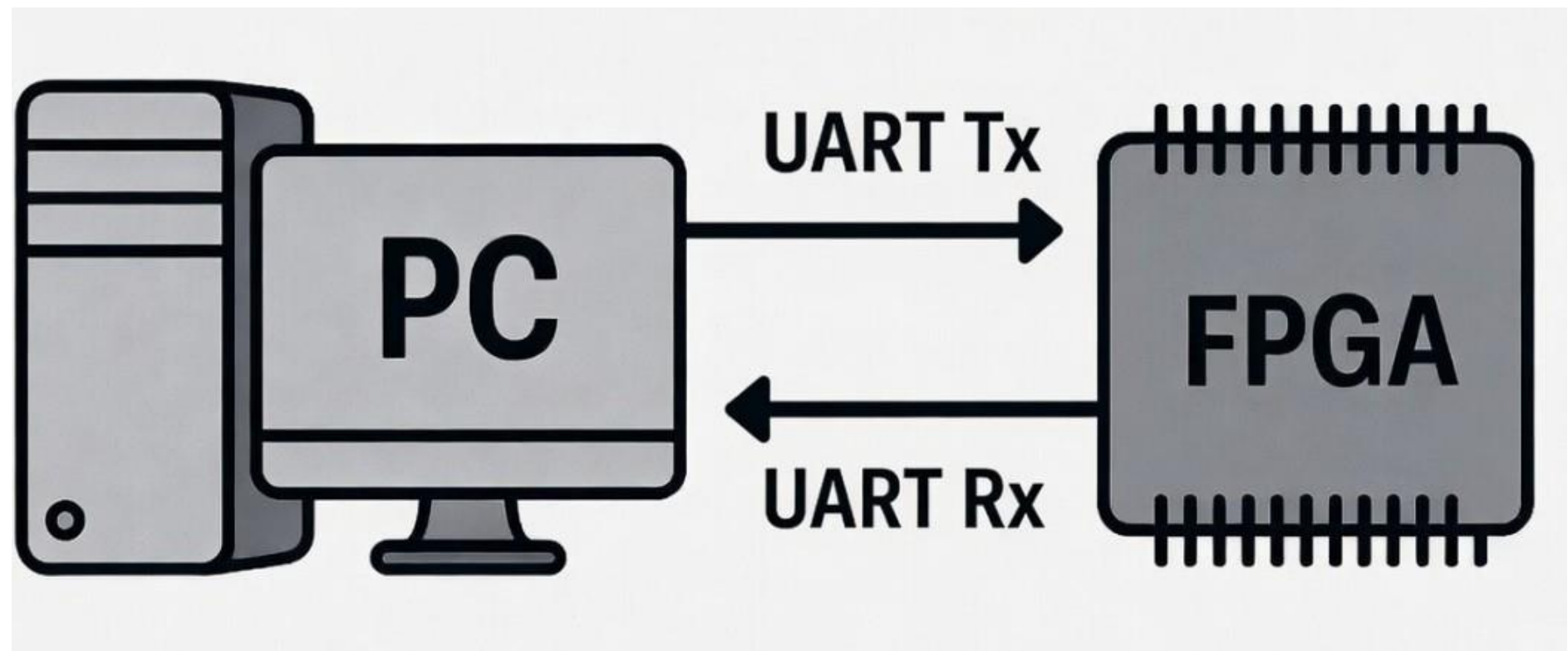

Figure 1: System Diagram

## 2.3 Design Flow

The design flow for this project was intentionally built from the core image-processing function outward. The first step was developing the image process core, because this module defines the main purpose of the system: scanning a 16 × 16 grayscale image, detecting offset or gain errors based on the selected mode, correcting defective pixels, and storing error locations. Starting with this module made it possible to define the memory format early, including the use of 8-bit grayscale pixel values, row-major pixel ordering, fixed threshold comparisons, and a reserved BRAM address region for storing detected error locations. Once the image processor behavior was defined, the Block Memory Generator IP was configured to match the required data path.

After the image-processing and memory structure were established, the UART receive and transmit paths were added. The UART RX module was designed to receive one byte at a time from the Python script, while a separate RX controller counted incoming bytes and wrote them sequentially into BRAM. This separation kept the serial communication logic independent from the image-storage logic. The UART TX module was then developed separately, followed by a TX controller that read corrected image bytes from BRAM and transmitted them back to the PC. This modular approach made the system easier to debug because each block had a clear role: UART RX/TX handled serial byte transfer, the RX/TX controllers handled BRAM addressing, and the image processor handled detection and correction.

The final step was integrating the modules in the top-level design using a control FSM. The top-level FSM sequenced the system through receiving the image, processing it, transmitting the corrected image, and then entering a display state. This order was chosen so that only one subsystem controlled the BRAM at a time, avoiding bus conflicts between the RX controller, image processor, TX controller, and LCD display logic. The LCD was added last as a bonus feature because the main pipeline could already demonstrate the project goal through Python image transfer, FPGA correction, returned image display, and LED error-count output. Adding the LCD at the end allowed the design to preserve the working data path while extending the system to read stored error locations from BRAM and display the selected offset or gain error information after processing was complete.

# 3 Detailed Hardware Design

## 3.1 System Hardware Architecture

As briefly explained in section 2.2, the main pipeline involves creating a test image in Python and sending to the FPGA through UART, from the UART the data is stored to the BRAM, then analyzed and corrected by the image processing core, stored back to the BRAM, then sent back to Python through the UART again. The FPGA uses a modular design where each block has one main responsibility. The UART modules handle serial byte transfer, the RX/TX controllers handle BRAM addressing, the image processor handles detection/correction, and the top-level FSM controls which module owns BRAM at each stage.

#### 3.1.1 Top-Level System Flow

Figure 2 shows the top-level system flow of the design. A clock wizard drives each module with a 10 MHz signal. At the center of the design is the BRAM where the image data is stored. Each module interacts with the BRAM and only one module is interacting at a time to avoid errors. First an image with errors is sent from the host PC to the FPGA through the UART Rx. The UART Rx includes a controller that handles where each received byte is stored on the BRAM. Once the entire image is stored to the BRAM, a signal is sent to start the image process module. This will read from the BRAM each byte and check for ”defective” pixels based on the pixel value. If there is an error, that error location is stored to the BRAM, and the pixel is corrected and overwritten inside the BRAM. After the image processing is done, a signal is sent to the UART Tx to start transmitting the corrected image back to the host PC. Again, the UART Tx includes a controller that reads the data from the BRAM before sending. Once the image is received, the number of errors is displayed on the FPGA’s LEDs. Finally, the LCD module reads the errors stored to the BRAM and displays them.

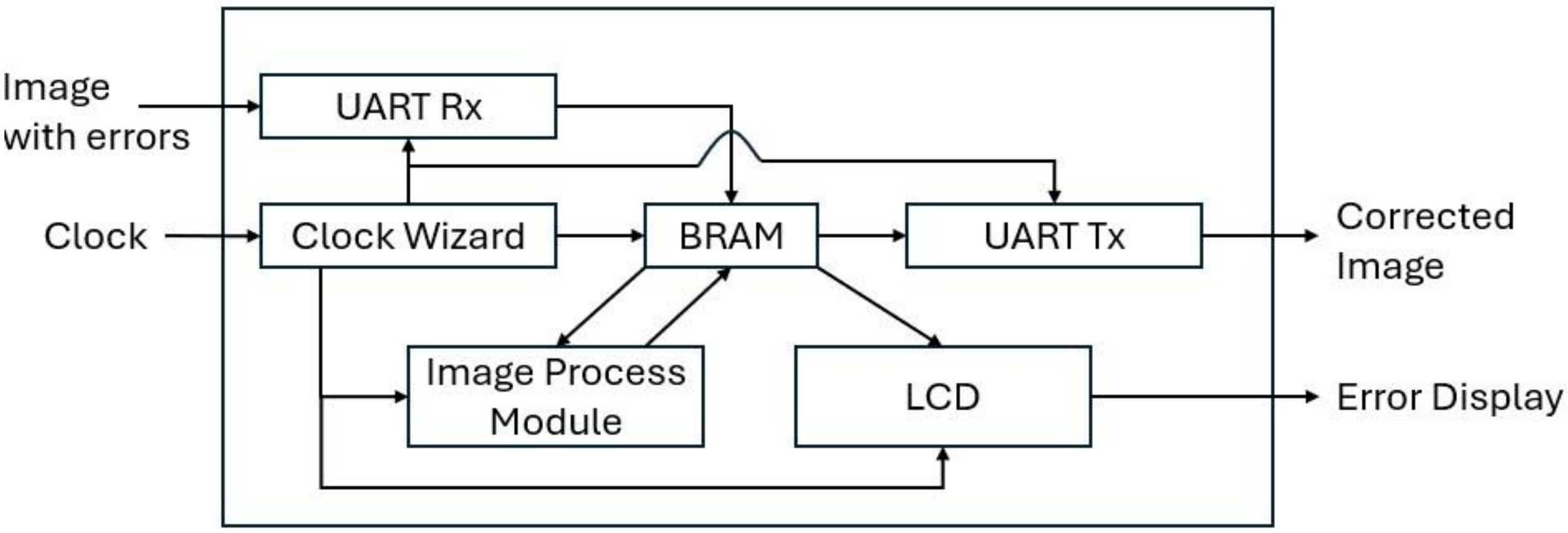

Figure 2: Block Diagram of Design

#### 3.1.2 Memory Architecture and BRAM Organization

The BRAM is the center of the architecture in this design. It is used as the frame buffer and error-list storage. The pixel data is 8-bit grayscale, so each BRAM word stores one pixel. For this design, a 16 x 16 image (either dark or light) is sent, so the first 256 addresses (0-255) of the BRAM are used to store the image. The addresses starting 256 store any detected error locations. Each error location is packed into one byte as {y,x}. Since x and y are both 4 bits, one 8-bit BRAM word can store one error coordinate. The design uses 9-bit BRAM addressing so both the image and error-list can fit in the same memory.

### 3.1.3 Image Processing Core

The image processor starts after the receive controller finishes storing the image in BRAM. For this design, the processing mode is selected by one DIP switch. When the switch is low, the module analyzes the image as a dark frame and checks for offset errors. When the switch is high, the module analyzes the image as a light frame and checks for gain errors. In dark mode, any pixel above the dark threshold is treated as an offset error. In light mode, any pixel above the high-gain threshold or below the low-gain threshold is treated as a gain error. Once an error is detected, the processor stores the error location in the reserved BRAM error-list region, increments the error counter, and writes a corrected value back into the image memory. For the dark image, the corrected value is the dark baseline of 10 grayscale counts. For the light image, the corrected value is the light baseline of 180 grayscale counts.

The image processing core was also verified through simulation using a behavioral BRAM model. In the zoomed-in waveform, the process start pulse begins the image-processing sequence, and the BRAM address increments sequentially through the image memory. This confirms that the image processor scans the image in row-major order, with each BRAM address corresponding to one 8-bit grayscale pixel. The simulation also shows bram dout providing the pixel value read from memory, while bram addr advances to the next pixel location after each processing cycle.
This verified that the core was reading through the stored image rather than randomly accessing memory. Figure 3 shows the zoomed-in waveform.

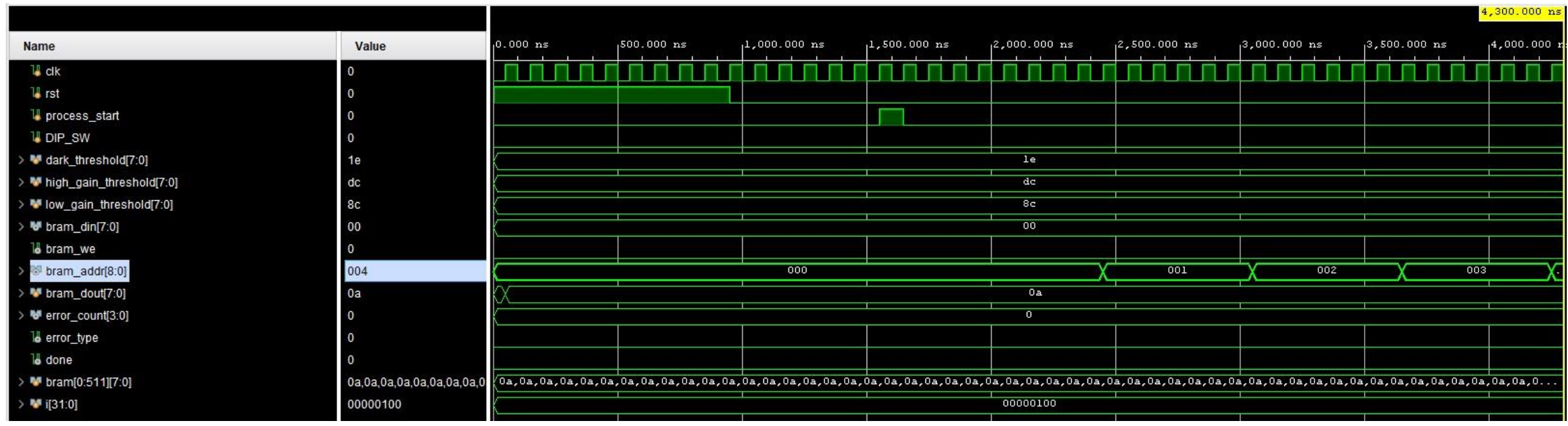

Figure 3: Image Process Core simulation

The zoomed-out waveform shows the error-detection and correction behavior during the dark-image test. When a pixel value exceeds the dark threshold, the module identifies it as an offset error. At each detected error, error count increments, bram we pulses, and bram _din is driven with the correction value or stored error-location data, depending on the current FSM state. The waveform shows the error count increasing as the processor reaches each deliberate offseterror pixel. This simulation was important because it verified the internal timing of the imageprocessing FSM before testing on hardware. In particular, it showed that the processor could step through the BRAM addresses, detect threshold violations, write correction data back to memory, and maintain a count of detected errors for later LED display. Figure 4 shows the zoomed-out waveform.

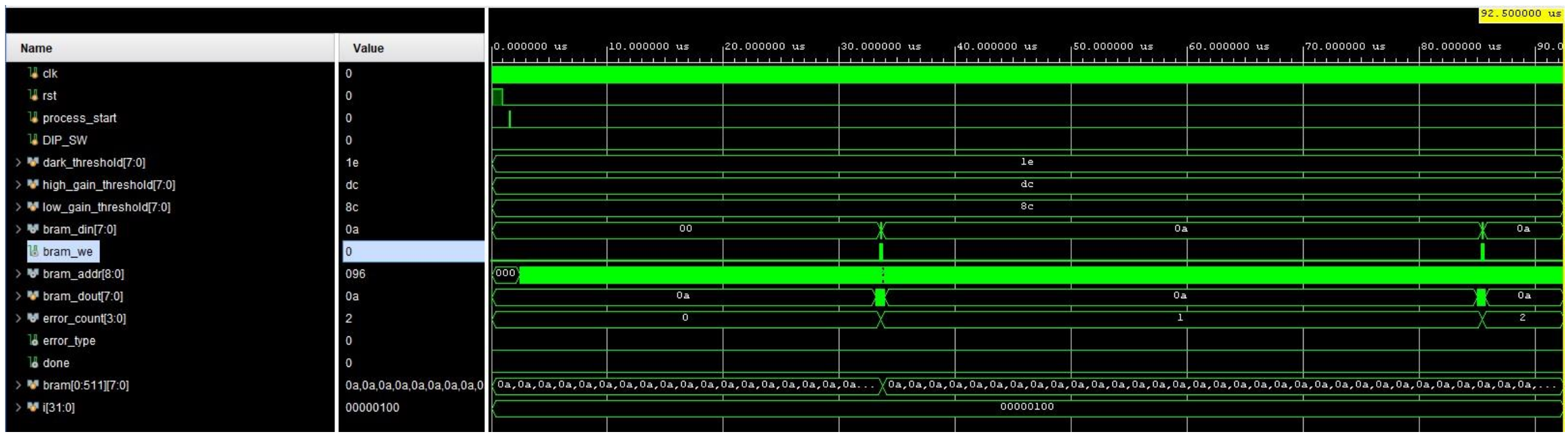

Figure 4: Image Process Core Simulation Expanded

### 3.1.4 UART Receive and Transmit

The UART RX and TX are implemented as separate modules, as well as the controllers for each. These modules don’t know what an image is, they only send and receive bytes. The controllers are what define the image length and BRAM address sequence. The RX module converts incoming serial data from the PC into 8-bit bytes. The RX controller then writes each received byte to the BRAM and counts until all 256 bytes (for 16 x 16 image) are received. The TX controller reads the pixels, including the new corrected ones, from the BRAM and sends them to the UART TX module. The UART TX module serializes each byte and sends it back to the PC. The design uses 9600 baud with a 10 MHz clock. Using one clock domain for the entire system avoided clockdomain crossing issues. The UART is also kept separate from image logic so each part could be debugged independently.

Two simulations were used to verify the UART communication path. The first simulation was a UART TX/RX loopback test, where the output of the UART TX module was connected directly to the input of the UART RX module. A test byte of 0x55 was loaded into the transmitter and sent using a tx start pulse. The transmitter asserted tx busy during the serial frame and pulsed tx _done when the byte was finished. The receiver then asserted rx done and reconstructed the same value, 0x55, on rx byte. This verified that the TX and RX modules used compatible baud timing, start and stop bit formatting, and LSB-first data ordering. Figure 5 shows the simulation of the UART TX/RX loopback test.

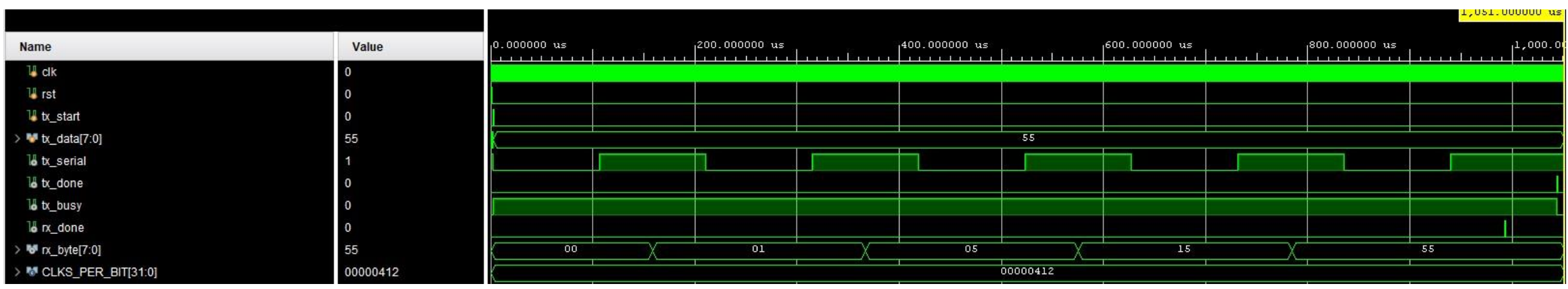

Figure 5: UART TX/RX loopback simulation

A second simulation verified the RX and TX controllers with a behavioral BRAM model. To reduce simulation time, a smaller 16-byte test transfer was used instead of the full 256-byte image. The RX controller was started with a start receive pulse, and each rx done pulse caused the next test byte to be written into BRAM. The write address incremented from 0x000 to 0x00F, and receive done asserted after the final byte was stored. The TX controller was then started with start transmit, read the same BRAM addresses sequentially, placed each stored value on tx data, and pulsed tx start for each transfer. The transmitted values matched the stored BRAM values, confirming that the controllers correctly handled byte counting, BRAM addressing, write-enable timing, and UART handshaking. These simulations were important because the UART path depends on precise timing: the bit-level UART timing must reconstruct bytes correctly, while the controller timing must align BRAM addresses and data so that the image is received and returned in the correct order. Figure 6 shows the simulation of the RX and TX controller behavior.

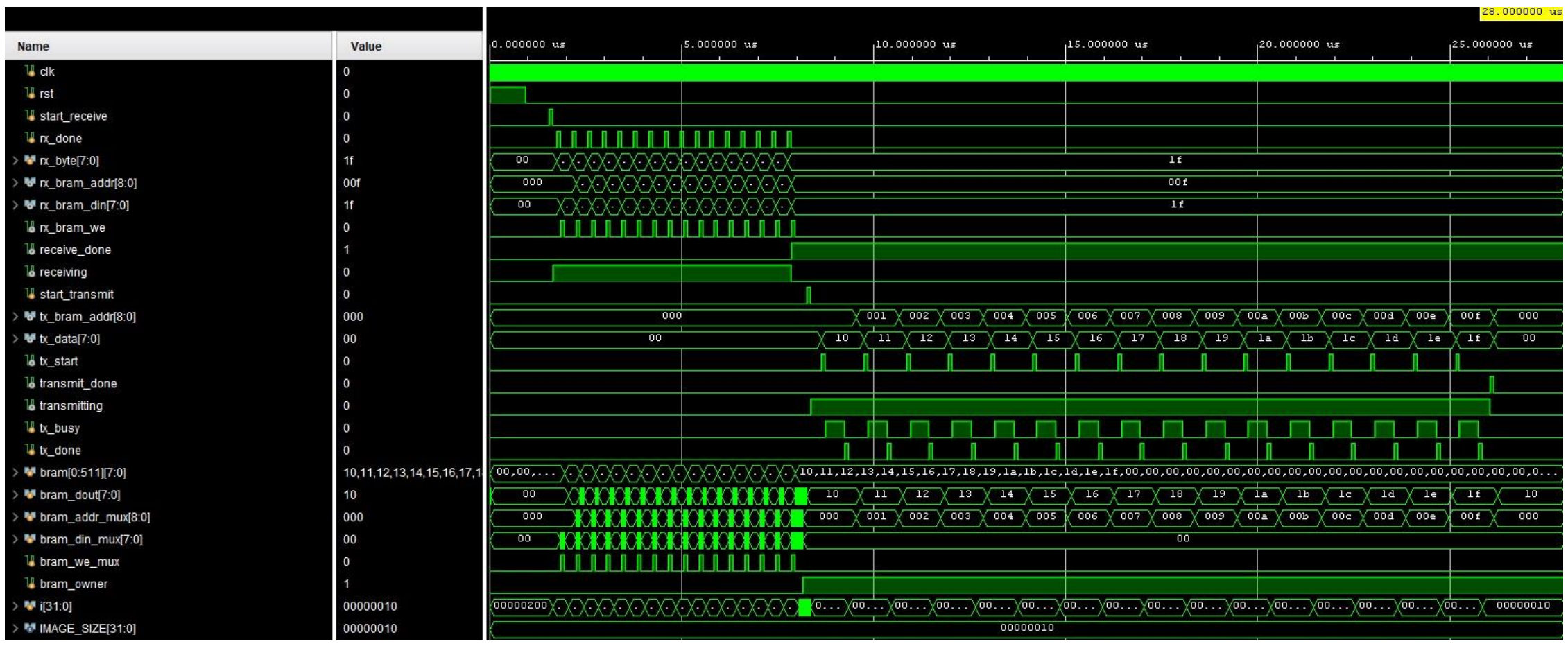

Figure 6: RX and TX controller simulation

### 3.1.5 Top-Level Control FSM and BRAM Arbitration

Since the design uses a single-port BRAM, only one subsystem can access the memory interface during a given clock cycle. The UART receive controller, image processing core, UART transmit controller, and LCD display logic all require BRAM access at different stages of the pipeline, but allowing them to drive the BRAM address, data, and write-enable signals simultaneously would create bus contention and unpredictable memory behavior. To avoid this, the top-level FSM controls BRAM ownership. BRAM muxing is used to select which module drives the address, write-enable, and data-in. During the receive state, the RX controller writes incoming image bytes to BRAM; during processing, the image core reads and updates pixel data; during transmission, the TX controller reads corrected pixels; and during display, the LCD logic reads the stored error list. After each subsystem is finished, a done signal is sent to the next subsystem to start it. This arbitration ensures deterministic memory access and keeps the modules independent while sharing the same memory resource.

### 3.1.6 LCD Error Display Interface

The LCD was added after the main pipeline as a display feature. After the corrected image is returned to the host PC, the FSM enters a display state. The LCD module reads the error locations stored on the BRAM and converts the stored {y,x} byte to ASCII characters for the display. The display button triggers the LCD to read and display the selected error. The left and right buttons will scroll through the stored errors. The display button must be pushed after each time the left or right button is pushed to display the coordinate. The center button will shift the display to read the whole message. In dark mode the first line will display ”OFFSET ERRORS:”, while in light mode the first line will display ”GAIN ERRORS:”. In either mode, the second line will display ”ERROR # (error index) of (error count), X = (x location), Y = (y location)”. If there are no errors, line 1 will display ”NO ERRORS” and line 2 will display ”IMAGE PASSED CHECK”. The system will leave the display state after reset is pulsed again and will then be ready for a new image to be sent. More details on the timing and structure of the LCD can be found in class notes created by Dr. Perera [5].

## 3.2 Peripherals Used

This design uses several board resources and some external resources. Python is used on the host PC to generate the test images, send the bytes, receive the corrected image, and display the result. The USB-to-UART bridge must obviously be used to send and receive images from the host PC. This project uses the Artix-7 AC701 Evaluation Board. The board has 365 36-Kb BRAM blocks, a small portion of it used for this project. The design uses one of the dip switches to toggle between dark mode or light mode. The LEDs on the board are used to display the number of errors found in the image. Since there are only 4 LEDs on the board, this design can only display a maximum of 16 errors. The LCD is used to display the type of error and the location of the error. Finally, Table 1 below summarizes the pushbuttons used and their role.

Table 1: Pushbutton Usage

| Pushbutton | Name | Purpose |
|---|---|---|
| NORTH | Reset | Resets FPGA |
| CENTER | Shift | Shifts LCD display |
| SOUTH | Display | Displays error location on LCD |
| WEST | Left | Cycles through the stored errors on BRAM to the left |
| EAST | Right | Cycles through the stored errors on BRAM to the right |

## 3.3 Problems Encountered and Solutions

There were a few problems encountered during the development of this project, some very minor and some more challenging. On the UART side, it is important that the correct CP210x driver is installed on the PC before Python can access the COM port. If the correct driver is not installed, Python will not be able to find the COM port even if the correct port is selected. Also on the UART side, it's important that the FPGA ports are connected correctly. The data input to the Rx module must be connected to the TX port, and the output data of the Tx module must be connected to the RX port in the constraints. This was originally flipped when the design was first implemented and Python was not able to send or receive the image correctly. Both of these were very simple fixes.

There were a couple more challenging problems in terms of the BRAM. There are several modules that need access to the BRAM to function properly, and the modules accessing the BRAM all at the same time could cause incorrect behavior, as stated earlier. As explained, this problem is solved by muxing the inputs and outputs to the BRAM so that only one module is controlling the BRAM at a time. The BRAM latency also presented some challenges. If the latency is not accounted for, the image process core will not read the pixels correctly and the ”defective” pixels will not be fixed. In order to ensure the BRAM latency is accounted for, several clock cycles were added between setting the BRAM address and reading the pixel value so that the correct pixel is read and is corrected. There is still some latency problems in the final design, which will be explained further in the next chapter.

Finally, there are some timing and display lags associated with the LCD. After scrolling through an error, the display sometimes requires an additional display press for the LCD to display the correct error location. This could be because the LCD needs to read the BRAM, capture the data, build the strings based on the stored data, and send the character across several FSM states.

# 4 Discussion and Conclusion

## 4.1 Discussion

The design was successfully implemented and programmed to the board with minor bugs still persisting. Overall, the design was successful in detecting the offset errors on the dark image, as well as both low-gain and high-gain errors on the light image.

### 4.1.1 Implementation Results

Figure 7 shows the implementation results of the design.

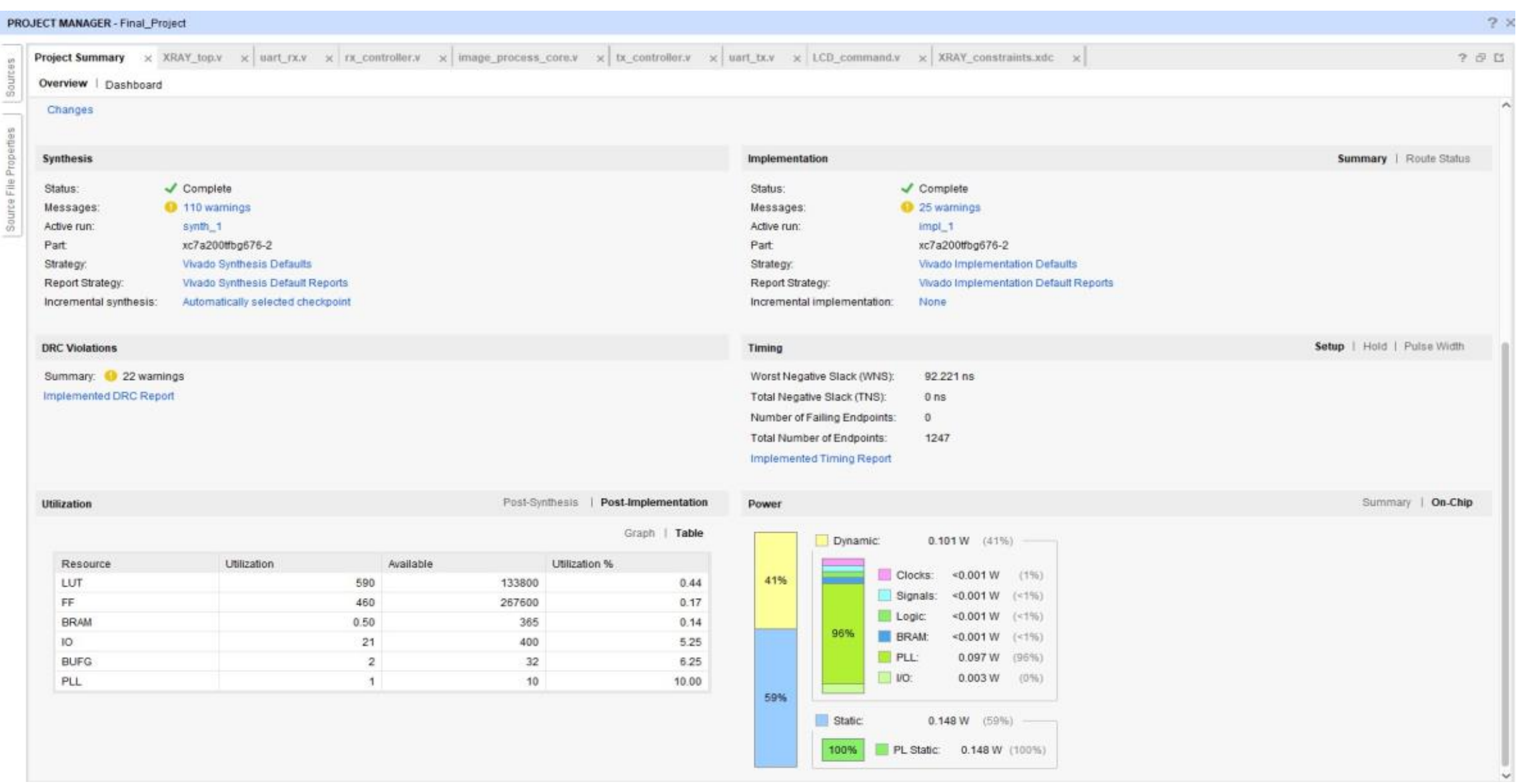

| Resource | Utilization | Available | Utilization % |
|---|---|---|---|
| LUT | 590 | 133800 | 0.44 |
| FF | 460 | 267600 | 0.17 |
| BRAM | 0.50 | 365 | 0.14 |
| IO | 21 | 400 | 5.25 |
| BUFG | 2 | 32 | 6.25 |
| PLL | 1 | 10 | 10.00 |

Figure 7: Implementation details

The implementation results show that the final design remains relatively lightweight even though it is more complex than the previous lab projects. The system includes UART receive and transmit modules, BRAM-based image storage, an image-processing FSM, LCD display control, button debouncing, and a top-level control FSM, but it still uses only 0.44% of the LUTs and 0.17% of the flip-flops. This low utilization suggests that the Artix-7 device has significant remaining capacity for future expansion. For example, the design could be extended to support larger image sizes, additional correction algorithms, larger error lists, or more advanced calibration methods without approaching the resource limits of the FPGA. Similarly, only half of one BRAM is used, so the current 16 × 16 image size is mainly a demonstration choice rather than a hardware limitation.

The timing results are also strong. The design meets timing with no violations, and the reported WNS of 92.192 ns shows a large timing margin relative to the 10 MHz system clock. This is especially encouraging because the implemented design contains 1273 timing endpoints, meaning the timing analyzer checked a fairly large number of register-to-register and I/O-related paths across the different modules. The large positive slack indicates that the FSM-based architecture, UART controllers, BRAM access logic, and LCD control path are all comfortably operating within the selected clock period. This also supports the decision to use a single 10 MHz clock for the system, since it simplified clock-domain management while still providing more than enough speed for the image-processing and communication tasks.

The power results show a total estimated power of 0.239 W, with most of the power coming from static power and clocking resources rather than the BRAM or logic itself. The slightly higher static power compared with previous labs, which were about 0.01 W lower, is reasonable because this project uses more of the board-level resources and includes additional I/O activity, UART communication, LCD outputs, and a larger integrated design. Since static power is dominated by the FPGA device being configured and powered rather than by the small amount of active logic used, the increase is not a major concern. The BRAM contribution remains very small, which further supports the possibility of increasing the memory usage for larger images in future versions without significantly increasing the total power.

#### 4.1.2 Dark Image Demo Results

For the dark image demo, a grayscale image was created in Python with low baseline pixel values of approximately 10 grayscale counts. The goal of this demo was to show the ability of the FPGA design to detect and correct pixels with offset error. In this mode, the DIP switch was set to select the dark-image analysis path, and any pixel above the selected dark threshold was identified as an offset error. The correction logic then replaced the detected offset-error pixels with the expected dark baseline value. Table 2 shows the locations of the deliberate errors added to the image. The X value refers to the column number, and the Y value refers to the row number. The LCD module converts the stored BRAM error locations into ASCII, so coordinate values from 10–15 are displayed as A–F.

Table 2: Offset Error Locations

| Error # | X | Y |
|---|---|---|
| 1 | 4 | 3 |
| 2 | A | 8 |
| 3 | 2 | C |

Figure 8 shows the dark image sent to the FPGA on the left and the corrected image returned to the PC on the right. The deliberately added high-offset pixels are removed in the returned image, showing that the threshold detection and correction path operated as intended for the darkframe case. The LEDs also displayed the correct number of errors, 3, and the LCD displayed the correct locations and error index for each stored location. For example, Error 1 was displayed at X = 4, Y = 3.

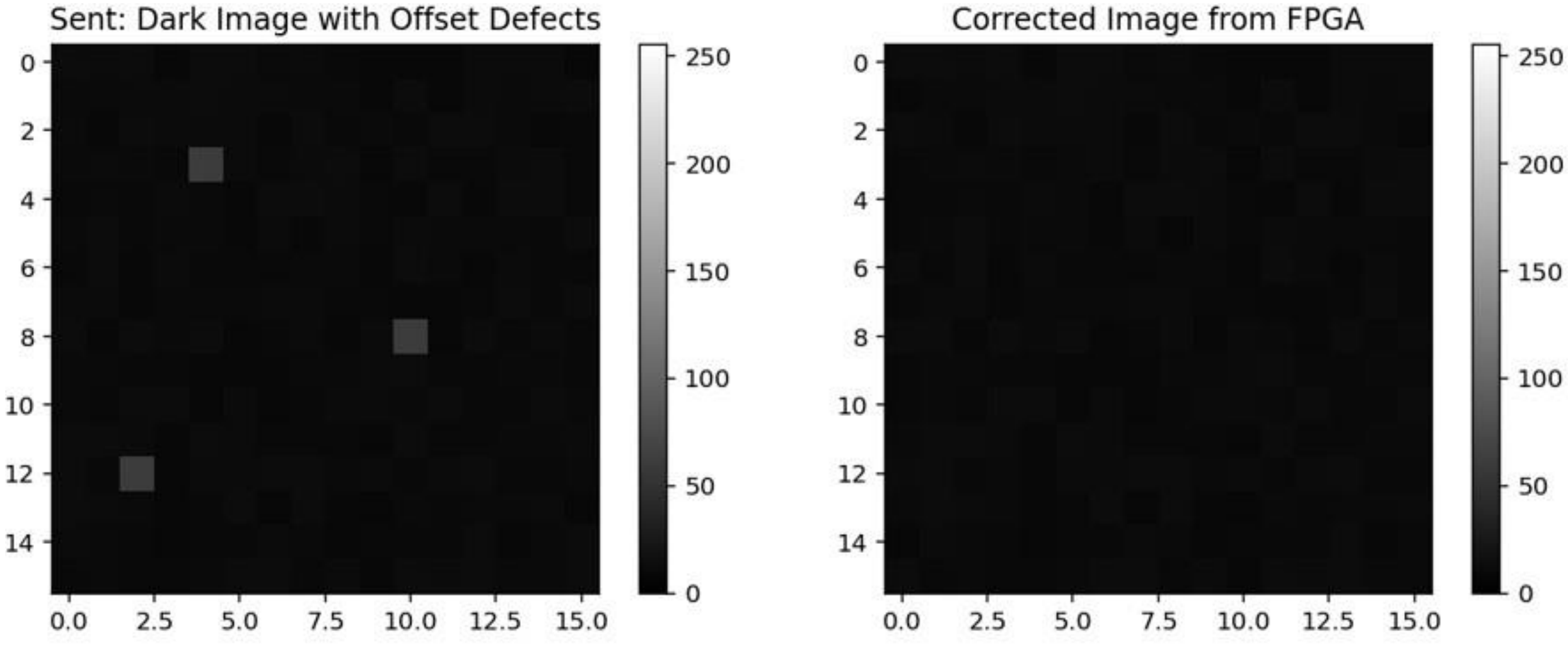

Figure 8: Dark Image Results: Left – image sent to FPGA; Right – corrected image

4.1.3 Light Image Demo Results

For the light image demo, a grayscale image was created in Python with pixel values randomized around 180 grayscale counts. The goal of this demo was to show the ability of the FPGA design to detect and correct pixels with gain error. In this mode, the DIP switch was set to select the light-image analysis path, and any pixel above the high-gain threshold or below the low-gain threshold was identified as a gain error. Like the dark image demo, the correction logic replaced the detected gain-error pixels with the expected light baseline value. Table 3 shows the locations of the deliberate errors added to the image.

Table 3: Gain Error Locations

| Error # | Error Type | X | Y |
|---|---|---|---|
| 1 | High gain | 5 | 5 |
| 2 | High gain | C | 9 |
| 3 | Low gain | E | 2 |
| 4 | Low gain | 7 | D |

Figure 9 shows the light image sent to the FPGA on the left and the corrected image returned to the PC on the right. The deliberately added high-gain and low-gain pixels are removed in the returned image, showing that the threshold detection and correction path operated as intended for the light-frame case. As before, the LEDs displayed the correct number of errors, 4, and the LCD displayed the correct locations and error index for each stored location. For example, Error 1 was displayed at X = 5, Y = 5.

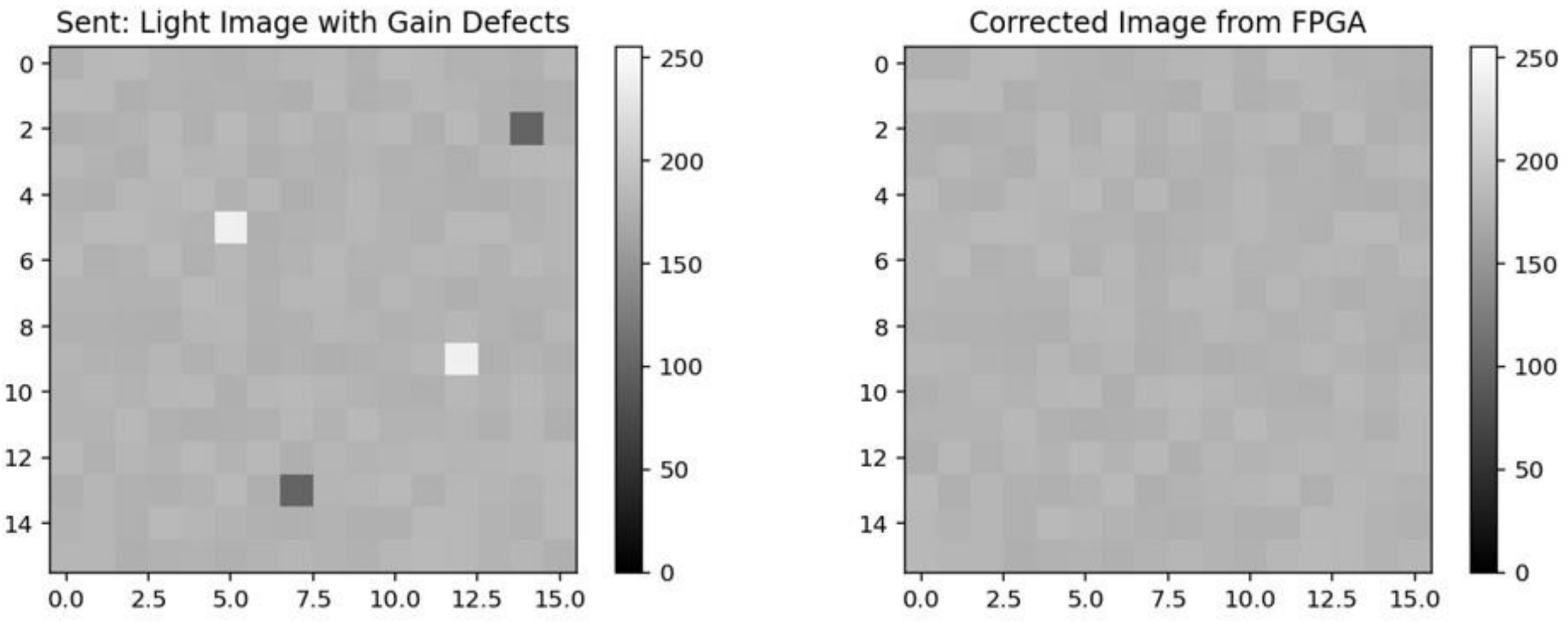

Figure 9: Light Image Results: Left – image sent to FPGA; Right – corrected image

Looking more closely at the corrected image, the returned image shows a one-column shift after correction. This is believed to be related to the BRAM latency issue discussed earlier in Section 3.3. Since BRAM reads are synchronous, the pixel data, address counter, write-back operation, and UART transmit path must remain aligned across multiple FSM states. A one-cycle mismatch in this sequence can cause corrected pixel data to be written to, or transmitted from, the neighboring address. Because the image is stored in row-major order, this appears visually as a one-column shift. Although this issue remains in the final prototype, the design still demonstrates the intended detection and correction behavior because the injected gain-error pixels are identified, corrected to the expected baseline value, counted correctly by the LEDs, and stored correctly for LCD display.

## 4.2 Future Work

Future work should first focus on resolving the remaining timing and synchronization bugs in the final prototype. The most important issue is the one-column shift observed in the returned corrected image. This is likely related to BRAM read latency or a one-cycle mismatch between the pixel address, captured pixel data, write-back address, and UART transmit sequence. A future version should more carefully pipeline these operations, possibly by adding explicit address/data registers or additional FSM states to guarantee that the corrected pixel is always written back to the intended memory address. The LCD display behavior could also be improved. Currently, after using the left or right button to change the selected error index, the display sometimes requires an additional display-button press before the correct location appears. This could be improved by automatically triggering a BRAM read and LCD refresh whenever the selected error index changes.

Another important improvement would be making the correction algorithm more representative of real detector calibration. The current design uses fixed thresholds and replaces defective pixels with known baseline values. This is acceptable for a controlled demonstration, but real image correction systems often use calibration frames and image statistics. A more realistic FPGA implementation could use a two-pass process, where the first pass calculates an average dark-frame or flat-field response and the second pass compares each pixel against that computed reference before applying correction. The gain correction could also be expanded from simple baseline replacement to a more realistic scaling operation based on a stored gain map or flat-field calibration factor.

The project could also be expanded to support larger images and additional defect types.

The implementation results show that the current design uses only a small amount of LUTs, flipflops, and BRAM, so there is significant room to increase the image size beyond 16 × 16. Future versions could use larger BRAM regions, multiple BRAMs, or streaming buffers to handle larger test images. Additional image-quality checks could also be added, such as dead pixels, stuck pixels, noisy pixels, bad rows or columns, or neighboring-pixel interpolation for correction. These improvements would make the system closer to a practical detector front-end while still preserving the modular FPGA architecture developed in this project.

## 4.3 Conclusion

The final project successfully demonstrated a simplified FPGA-based mini X-ray detector front-end system. The completed design was able to receive a 16 × 16 grayscale image from Python through UART, store the image in BRAM, analyze the image for offset or gain-related pixel errors, correct detected error pixels, return the corrected image to the PC, display the number of detected errors on the LEDs, and display stored error locations on the LCD. The dark image demo correctly detected and corrected the deliberate offset errors, while the light image demo correctly detected both high-gain and low-gain errors. In both cases, the LED count and LCD error-location display matched the expected number and location of the inserted errors.

Overall, the original goal of the project was achieved. The design did not attempt to reproduce a complete commercial X-ray detector front end, but it did demonstrate the major digital processing steps that such a front-end system would require: image transfer, frame-buffer storage, error analysis, correction, and output communication. The project also showed the usefulness of a modular FPGA design approach. By separating the UART modules, BRAM controllers, image processing core, LCD interface, and top-level FSM, each part of the system could be developed and debugged independently before being integrated. Although minor bugs remain, especially the one-column shift in the returned light image and occasional LCD display lag, the final implementation met timing, used only a small fraction of the FPGA resources, and successfully demonstrated the intended proof-of-concept pipeline.

This work is inspired by the digital design research group at UCCS. This group has done extensive work in FPGA-based architectures, techniques, and associated models. Their analyses [6],[7] show that FPGA-based embedded systems are currently the best option to support applications and techniques, such as the ones presented in this report. Also, their previous work on FPGA-based embedded accelerators, architectures, and techniques for various compute and data-intensive applications, including data analytics/mining [8],[9],[10],[11],[12],[13],[14],[15],[16],[17]; control systems [18],[19],[20],[21],[22],[23]; cybersecurity [24],[25],[26]; machine learning [27],[28],[29],[30],[31],[32],[33],[34]; communications [35],[36]; edge computing [37],[38],[39]; bioinformatics [40],[41]; and neuromorphic computing [42],[43],[44]; demonstrated that FPGA-based embedded systems are the best avenue to support and accelerate complex algorithms and techniques.

Also as future work, we are planning to investigate FPGA-based hardware optimization techniques, such as parallel processing architectures (similar to [29],[45],[46],[47]), partial and dynamic reconfiguration traits (as stated in [48],[49],[50]) and architectures (similar to [25],[51],[52],[53]), HDL code optimization techniques (as stated in [54],[55]), and multi-ported memory architectures (similar to [56],[57],[58],[59]), to further enhance the performance metrics of FPGA-based embedded architectures for mini X-ray detector front end, while considering the associated tradeoffs.